\documentclass[11pt,letterpaper]{llncs}

\usepackage{multirow}
\usepackage{amstext}
\usepackage{amssymb}
\usepackage{mathrsfs}
\usepackage{array}
\usepackage{subfigure}

\usepackage{algorithm}
\usepackage{algorithmicx}
\usepackage{algpseudocode}

\usepackage{hyperref}

\usepackage{epsfig}
\usepackage{epstopdf}
\usepackage{pdftricks}
\usepackage{algorithm}
\usepackage{algorithmicx}

\newtheorem{thm}[theorem]{Theorem}
\newtheorem{lem}[theorem]{Lemma}
\newtheorem{cor}[theorem]{Corollary}
\newtheorem{defn}[theorem]{Definition}
\newtheorem{observation}[theorem]{Observation}
\newcommand{\A}{\mathcal{A}}
\newcommand{\W}{\mathcal{W}}
\newcommand{\set}[1]{\left\{#1\right\}}

\begin{document}

\title{Envy, Multi Envy, and Revenue Maximization}

\author{Amos Fiat and Amiram Wingarten}
\institute{School of Computer Science, Tel Aviv University}

\maketitle


\begin{abstract}
We study the envy free pricing problem faced by a seller who wishes to maximize revenue by setting prices for bundles of items.
Consistent with standard usage \cite{DUB61} \cite{FOL67}, we define an allocation/pricing to be \textit{envy free} if no agent
wishes to replace her allocation (and pricing) with those of another agent.

If there is an unlimited supply of items and agents are single minded then we show that finding the revenue maximizing envy free
allocation/pricing can be solved in polynomial time by reducing it to an instance of weighted independent set on a perfect graph.

We also consider a generalization of envy freeness. We define an allocation/pricing as \textit{multi envy free} if no agent wishes
to replace her allocation with the union of the allocations of some set of other agents and her price with the sum of their prices.
We show that even though such allocation/pricing can be approximated by $O(\log m+\log n)$ factor ~\cite{Bal08}, it is \textit{coNP}-hard
to decide if a given allocation/pricing is multi envy free. We also show that revenue maximization multi envy free allocation/pricing
is \textit{APX} hard.

An interesting restricted version of the subset pricing problem is when all items are intervals of a line segment and all requests are
a contiguous set of items along the line. The motivation here is that one can think of the agents as drivers on a highway when each
product is highway segment(or guests in a hotel --- items are translated to dates). In this setting, determining if a given allocation/pricing
is multi envy free is polynomial time. If the highway has bounded capacities then a revenue maximizing envy free allocation/pricing can be computed in
polynomial time and we also give an FPTAS for the revenue maximizing multi envy free allocation/pricing.
\end{abstract}


\section{Introduction}

We consider the combinatorial auction setting where there are several different items for sale, not all items are identical, and agents have valuations for subsets of items. We allow the seller to have identical copies of an item. We distinguish between the case of limited supply (e.g., physical goods) and that of unlimited supply (e.g., digital goods).
Agents have known valuations for subsets of items. We assume free disposal, {\sl i.e.}, the valuation of a superset is $\geq$ the valuation of a subset.
Let $S$ be a set of items, agent $i$ has valuation $v_i(S)$ for set $S$.
The valuation functions, $v_i$, are public knowledge. Ergo, we are not concerned with issues of truthfulness or incentive compatible bidding. Our concern here is to maximize revenue in an envy free manner.

Our goal is to determine prices for sets of items while (approximately) maximizing revenue.
The output of the mechanism is a payment function
$p$ that assigns prices to sets of items and an allocation $a$. Although there are exponentially many such sets, we will only consider payments functions that have a concise representation. For a set of items $S$ let $p(S)$ be the payment required for set $S$. Let $a_i$ be the set assigned to agent $i$.

In general, every agent $i$ has valuation function $v_i$ defined over every subset of items.

Given a payment function $p$, and a set of valuation functions $v_i$, let
$z_i = \max_S (v_i(S)-p(S))$, and let $\mathcal{S}_i$ to be a collection of sets such that $S \in \mathcal{S}_i$ if and only if $v_i(S)-p(S) = z_i$.

We now distinguish between two notions of envy freeness.

\begin{defn}
We say that $(a,p)$ are envy free if
\begin{itemize}
\item If $z_i>0$ then $a_i \in \mathcal{S}_i$.
\item If $z_i=0$ then either $a_i \in \mathcal{S}_i$ or $a_i = \emptyset$.
\item If $z_i<0$ then $a_i=\emptyset$.
\end{itemize}
\end{defn}

\begin{defn}
If $a_i=\emptyset$ then we say that agent $i$ loses, otherwise we say that agent $i$ wins.
\end{defn}

\begin{defn}
A pricing $p$ is monotone if for each subset $S$ and for each collection of subsets $C$ such that $S \subseteq \bigcup_{T\in C}T$ the following inequality holds: $p(S) \leq \sum_{T\in C}p(T)$.
\end{defn}

\begin{defn}
An allocation/pricing $(a,p)$ is multi envy free if it is envy free and its pricing is monotone.
\end{defn}

Clearly, multi envy-freeness is a more demanding requirement than envy-freeness, so any allocation/pricing that is multi envy-free is also envy-free.
In item pricing, a price is set for every item, identical copies of an item are priced the same, and the price of a set is the sum of the individual item prices. In subset pricing one may assign a sets of items prices that cannot be consistently expressed as a sum of the item prices comprising the set. E.g., discounts for volume would not generally be consistent with item pricing.

In the unlimited supply setting, item pricing is always multi envy-free (and hence also envy-free).
With a limited supply of items, achieving item-pricing [multi] envy-freeness is not automatic. Circumstances may arise where some agent has a valuation less than the price of some set of items she is interested in, but there is insufficient supply of these items. An envy-free solution must avoid such scenarios.
Even so, for limited or unlimited supply, item pricing is envy-free if and only if item pricing is multi envy-free
(this follows from the monotonicity of the item pricing).

For subset pricing, it does not necessarily follow that every allocation/pricing that is envy-free is also multi envy-free.

Although the definitions above are valid in general, we are interested in single minded bidders, and more specifically in a special case of
single minded bidders called the highway problem (\cite{Gur05,Blu06}) where items are ordered and agents bid for consecutive interval of items.


\section{Our Results}

Table \ref{p:TAB_diffs} gives gaps in revenue between the item pricing (where envy freeness and multi envy freeness are equivalent), multi envy freeness, and envy freeness.
These gaps are for single minded bidders, and the gaps between item pricing and multi envy free subset pricing are in the context of the highway problem.
In all cases (single minded bidders or not) \begin{eqnarray*} \mbox{Revenue([Multi] EF item pricing)} &\leq& \mbox{Revenue(Multi EF subset pricing)} \\
&\leq& \mbox{Revenue(EF subset pricing)} \\
&\leq& \mbox{Social Welfare}.\end{eqnarray*}
Clearly, if a lower bound holds for unlimited supply it clearly also holds for (big enough) limited supply.

  All of our lower bound constructions are for single minded bidders,
  for single minded bidders with unlimited supply the bounds are almost tight as
  $$(\mbox{\rm Social welfare})/(\mbox{\rm Envy-free item pricing}) \leq H_m + H_n.$$ This follows directly from \cite{Gur05}, see below.

\begin{table}[t]
  \centering
\begin{tabular}{|c|c|c|c|c|c|c|}
  \hline
 & Lower & [Multi] Envy-free & Multi envy-free  & Envy-free & Type of \\
 & Bound  & item pricing & subset pricing & subset pricing & Instance \\
  \hline
Limited & $\#$1 & $H_n$ & $n$ & &Highway \\
        & & $H_m$ & $m$ &  &\\
  \hline
Unlimited & $\#$2 &  & $1$ & $H_n$ & Single Minded  \\
         &   &  & $1$ & $H_m$ & \\ \hline
Unlimited & $\#$3 & $1$ & $H_n$ & & Highway  \\
                    &   & $1$ & $\log\log m$ & &  \\
\hline
\end{tabular}
  \caption{Revenue gaps for single minded bidders ($n$ - $\#$ items, $m$ - $\#$ agents). }
  \label{p:TAB_diffs}
\end{table}

For limited supply, lower bound $\#$ 1 shows for some inputs the revenue of [Multi] Envy-free item pricing can be significantly smaller (by a factor
$\leq H_n/n$ or $\leq H_m/m$) than the revenue of Multi envy-free subset pricing. This gap is smaller for unlimited supply, lower bound $\#$ 3 shows that for unlimited supply it
is possible to achieve a ratio
of $1/H_n$ or $1/\log \log m$ between the
revenue of [Multi] Envy-free item pricing and that of Multi envy free subset pricing. Both lower bounds $\#$ 1 and $\#$ 3 are for the highway problem.

Lower bound $\#$ 2 in Table \ref{p:TAB_diffs} shows a gap in revenue ($1/H_n$ or $1/H_m$) between Multi envy-free subset pricing and Envy-free subset pricing. This bound is for single minded bidders, but not for the highway problem.

We further give several hardness results and several positive algorithmic results:

\begin{enumerate}
\item For unlimited supply, and single minded bidders, we show that finding the envy free allocation/pricing that maximizes revenue is polynomial time.
\item
We show that the decision problem of whether an allocation/pricing is multi envy free is \textit{coNP}-hard.
\item  We also show that finding an allocation/pricing
that is multi envy free and maximizes the revenue is \textit{APX}-hard.
\item
For the the highway problem, we show a that if
all capacities are $O(1)$ then the (exact) revenue maximizing {\sl envy free}
allocation/pricing can be computed in polynomial time. {\sl I.e.}, the problem is fixed parameter tractable with respect to
the capacity.
\item Again, for the highway problem with $O(1)$ capacities, we give a FPTAS for revenue maximization under the more difficult Multi envy-free requirements.
\end{enumerate}

\section{Related Work}

Much of the work on envy free revenue maximization is on item pricing rather than on subset pricing.
Guruswami et al.~\cite{Gur05} give an $O(\log m+\log n)$-approximation for the general single minded  problem, where $n$ is the number of items and $m$ is the number of agents. This result was extended by Balcan et al. ~\cite{Bal08} to an $O(\log m+\log n)$-approximation for arbitrary valuations and unlimited supply using single fixed pricing which is basically pricing all bundles with the same price.
Demaine et al. ~\cite{Dem06} show that the general item pricing problem with unlimited availability of items is hard to approximate within a (semi-)logarithmic factor.

The general combinatorial auction setting allows agents to request any arbitrary set of items. Even for unlimited supply, the problem of maximizing the revenue is hard to approximate within a (semi-)logarithmic factor ~\cite{Dem06} (and shown to be APX hard for a restricted case called the ``graph vertex problem" where agents are interested in sets of one or two items only ~\cite{Gur05}).

If every agent has an associated set of items such that the agent values any nonempty subset equally, ~\cite{Gur05} present a $O(\log n)$ approximation algorithm. Briest and Krysta ~\cite{Bri07} show that the revenue maximization problem is inapproximable within $O(\log^{\epsilon} m)$ for some $\epsilon > 0$. This indicates that we can not expect to do better in the more general setting where  valuations are for bundles and agents are not single minded.

For the limited supply general problem it is easy to show that it is NP-hard to approximate it within $m$ or $n^{1/2}$ as noted by Cheung and Swamy in ~\cite{Swam08} (Grigoriev et al. ~\cite{Gri06} show that it is NP-complete to approximate the maximum profit within a factor $m^{1-\epsilon}$ , for any $\epsilon > 0$ even when the underlying graph is a  grid). \cite{Swam08} gives an  $O(\sqrt{n}\log u_{max})$ approximation for envy free profit maximization problem (where $u_{max}$ = max item supply).

\noindent{\bf The Highway Problem}

For the unlimited supply the problem is NP-hard (Briest and Kriesta ~\cite{Kry06}). Also the problem is given $O(\log n)$-approximation by ~\cite{Blu06}. When the  length of each interval is bounded by a constant or the valuation of each agent is bounded by a constant, Guruswami et al. ~\cite{Gur05} give a fully polynomial time approximation scheme (FPTAS). If the intervals requested by different agents have a nested structure then an FPTAS is possible  ~\cite{Blu06,Kry06}.

For limited supply when the number of available copies per item is bounded by $C$, Grigoriev et al. ~\cite{Gri06} propose a dynamic programming algorithm that computes the optimal solution in time $O(n^{2C}B^{2C}m)$, where $n$ is the number of agents, $m$ the number of items, and $B$ an upper bound on the valuations. For $C$ constant, and by appropriately discretizing $B$, this algorithm can be used to derive an FPTAS for this version of the highway problem. However, the solution produced by this algorithm need not be envy-free. For the highway problem with uniform capacities, ~\cite{Swam08} gives an $O(\log u_{max})$ approximation algorithm where all capacities are equal, this algorithm does produce an envy-free allocation/pricing.


\section{Notation and Definitions}
In our setting we have $m$ agents and $n$ items.

The capacity of an item is the number of (identical) copies of the item available for sale. The supply can be \textit{unlimited supply} or \textit{limited supply}. In a \textit{limited supply} seller is allowed to sell up to some fixed amount of copies of each item. In the \textit{unlimited supply} setting, there is no limit as to how many units of an item can be sold.

We consider \textit{single-minded} bidders, where each agent has a valuation for a bundle of items, $S_i$,
and has valuation $0$ for all sets $S$ that are not supersets of $S_i$.
The valuation function for $i$, $v_i$, has a succinct representation as $(S_i, v_i)$ where $v_i = v_i(S_i)$. For every $S$ such that $S_i \subseteq S$, $v_i(S)= v_i(S_i) = v_i$, for all other sets $S'$, $v_i(S')=0$.

If $a_i=S$ and $S_i \subset S$, we can change the allocation to be $a_i=S_i$ and keep the same price. The original allocation/pricing is envy free if and only if the modified allocation is also envy free. Therefore, we can say without lost of generality, that an allocation/pricing must either have $a_i = \emptyset$ and $p_i(a_i)=0$ or $a_i=S_i$ and $p_i(a_i) \leq v_i(S_i)$.

We denote the set of agents $i$ allocated $S_i$ (the winning agents) $W = \{ i : a_i \neq \emptyset \}$ to be the set of agents for which $a_i = S_i$.
Our goal is to find an allocation/pricing that maximizes
$$\sum_{i \in W }p(S_i).$$ For single minded bidders, we say that agent $i$ wins if $i \in W$. Otherwise, we say that $i$ loses.

Fix the price function $p$.
For unlimited supply,it is easy to see that the revenue maximizing winner set $W = \{i:p({S_i}) \leq v({S_i})\}$.
For limited supply,  it must be that $$\{i:p({S_i})<v({S_i})\}\subseteq W \subseteq \{i:p({S_i})\leq v({S_i})\}.$$

\subsection{Envy and Multi Envy for Single minded bidders}

For single minded bidders, the definitions of envy free and multi envy free can be simplified as follows:
\begin{observation}
For single minded bidders an allocation/pricing is envy free if and only if
 \begin{enumerate}
   \item For any two winning agents $i$ and $j$, if $S_i\subseteq S_j$, it holds that $p(S_i) \leq p(S_j)$
   \item For each losing agent $i$ and winning agent $j$ such that $S_i\subseteq S_j$ it holds that $v({S_i}) \leq p({S_j})$
 \end{enumerate}
\end{observation} \label{p:OBS_envy free}
\begin{observation}
For single minded bidders an allocation/pricing is multi envy free if and only if
 \begin{enumerate}
   \item  For any winning agent $i$ and any collection of winning agents $C$ such that $S_i \subseteq \bigcup_{j \in C} S_j$ the following must hold: $p({S_i}) \leq \sum_{j \in C}p({S_j})$
   \item  For any losing agent $i$ and any collection of winning agents $C$ such that $S_i \subseteq \bigcup_{j \in C} S_j$ the following must hold: $v({S_i}) \leq \sum_{j \in C}p({S_j})$.
 \end{enumerate}
\end{observation}

\section{Revenue Gaps Between Models}

In this section we show the gaps between the optimal solutions of the different models.
It is clear that item pricing envy free setting is less profitable than the subset pricing multi envy free setting since any item pricing solution is also multi envy free solution. We give two theorems that show the gaps between the two models when items are given in limited supply and when items are given in unlimited supply.

The following theorem corresponds to line $\# 1$ in Table \ref{p:TAB_diffs}.
\begin{thm}
The maximal revenue in an envy-free item pricing for the limited supply highway setting may be as low as $\frac{H_k}{k}$ times the maximal revenue achievable by a subset pricing that is multi envy free(where $H_k$ is the $k$'th Harmonic number, and $k$ is the maximal capacity of an item).
\end{thm}

\begin{proof}
Consider a path of $k$ segments where segment $1$ has capacity of $k$ and for $i>1$, segment $i$ has capacity $n - i + 1$. There are two groups of agents in the setting:
\begin{itemize}
  \item  $k$ agents: $\set{ \text{ request } [1,i] \text{ with valuation of } 1 \,\,|\,\,\, 1\leq i \leq k}$
  \item $k-1$ agents: $\set{ \text{ request }  [i,i] \text{ with valuation of }  \frac{1}{i} \,\,|\,\,\, 1\leq i \leq k}$
\end{itemize}
See Figure \ref{fig3} for illustration.
\begin{figure} [h]
 \centering
 \scalebox{0.55}{\input{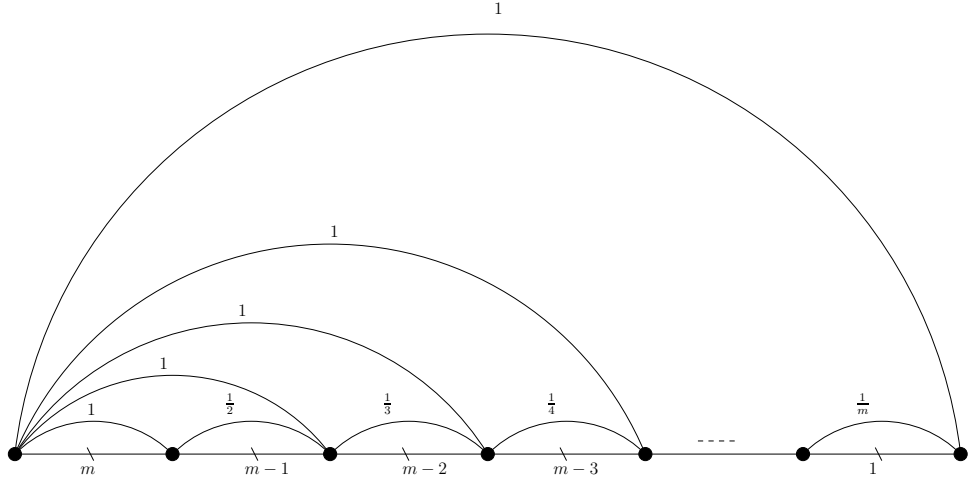}}
 \caption{The most profitable agents are the the one with valuation $1$, however allocating them with their bundles makes the others envy }
 \label{fig3}
 \end{figure}

Clearly the best an item pricing envy-free solution can achieve is ${H_k}$ by assigning price of $\frac{1}{i}$ to segment $i$.
A subset pricing multi envy free solution can price all intervals on the path at $1$ and get a revenue  of $k$.

The number of requests is $O(k)$, therefore the gap is $O(m/H_m)$.
\end{proof}

We now deal with the gaps between item pricing setting, multi envy free setting and envy free setting when items are
given in unlimited supply. It is clear that any item pricing solution is also multi envy free solution and that any multi envy
free solution is also an envy free solution. The question is how big can the gap be, how more can a seller profit by choosing
another concept of envy? Guruswami et al.~\cite{Gur05} give an $O(\log m+\log n)$-approximation for the general item
pricing problem. The upper bound used is the sum of all valuations. Since any item pricing solution is also multi envy free and
since the sum of all valuations can be used as an upper bound on multi envy free and envy free solutions, we conclude that
the gap between any pair of these problems is upper bounded by $O(\log m+\log n)$. We show lower bounds on the gaps. The following theorem corresponds to line $\# 3$ in Table \ref{p:TAB_diffs}.

\begin{thm}
Multi envy free subset pricing setting for the unlimited supply problem (even for a highway) can improve the optimal
item pricing solution by a factor of $O(\log\log m)$ where $m$ is the number of requests.
\end{thm}
\begin{proof}
We construct the following agent requests along a path. We assume for simplicity that $n$, the number of items, is a power of $2$, so $n = 2^k$.
We  have $k+1$ layers of requests, starting from layer $0$ up to layer $k$. In layer $i$ there are $2^i$ equal sized requests,
each for $2^{k-i}$ items, of valuation $\frac{1}{2^i(k-i)}$. See Figure \ref{fig6}. The multi envy-free pricing can accept all requests and achieve  revenue $O({H_k})$ while item pricing can produce at most $O(1)$ revenue. Since $k = O(\log m)$ this
 shows that the ratio between [multi] envy-free  item pricing revenue and multi envy free subset pricing revenue can be as low as  $O(1/\log\log m)$.
\begin{figure} [h]
 \centering
 \scalebox{0.55}{\input{Draw6.pstex_t}}
 \caption{ }
 \label{fig6}
\end{figure}

\end{proof}

The following theorem corresponds to line $\#2$ in Table \ref{p:TAB_diffs}.
\begin{thm}
 Envy free subset pricing setting for the unlimited supply problem can improve on the optimal multi envy free allocation/pricing
 solution by a factor of $O(\log m)$ where $m$ is the number of requests.
\end{thm}
\begin{proof}
We construct an instance with $m$ agents and $m$ items. The set of all items is denoted $V = \{1,...,m\}$.
For agent $i$ we have $S_i = V \setminus \{i\}$ with valuation $v_i = \frac{1}{i}$. Since every request is a subset of the union of any two other
requests, the minimal $v_i + v_j$ where $v_i$ and $v_j$ both, is an upper bound on the revenue that can be achieved from any allocated request in the
multi envy free setting. Let the two lowest winning valuations be  $i_\alpha$ and $i_\beta$.
All requests with lower valuations are not allocated. All requests with valuation higher than $i_\alpha$ and $i_\beta$
can have price that is at most $\frac{1}{i_\alpha} + \frac{1}{i_\beta}$. Therefore no multi envy free allocation/pricing
can achieve more than revenue $2$.

On the other hand, in the envy free setting, an allocation of each request at price $p_i=v_i$ is valid and achieves revenue of $H_m$.
Therefore, the ratio between maximal revenue achieved in the multi envy free setting {\sl vs.} the maximal revenue achieved in the envy free setting
can be as low as $O(1/\log m)$.
\end{proof}

\section{Polynomial time envy-free revenue maximization (unlimited supply, single minded bidders)}

We discuss in this section the classical envy free approach where each agent may envy only one other agent (not a group of other agents).
Therefore we seek an allocation/pricing such that Observation \ref{p:OBS_envy free} is valid.

With limited supply of items, a reduction from independent set can be used to show that the problem of maximizing revenue is hard to
approximate within $m^{1-\epsilon}$. Grigoriev et al. ~\cite{Gri06} show that it is NP-complete to approximate the maximum profit
limited supply item pricing solution to within a factor $m^{1-\epsilon}$ , for any $\epsilon > 0$ even when the underlying graph is a  grid.
The same construction can be used here.

For the unlimited supply setting we show that:
\begin{thm} \label{p:THM_EF}
 For single minded bidders the revenue maximizing envy free allocation/pricing with
 unlimited supply can be computed in polynomial time.
\end{thm}
Our idea is to make use of the fact that allocating a certain request at price $p$ means that any request that is a superset and has
valuation $< p$ must not be allocated.
We transform all requests into a directed acyclic perfect graph $H$ and then compute the revenue maximizing allocation/pricing by
computing a maximal independent set on $H$ (which can be done in polynomial time for perfect graph). A similar construction using
the hierarchical nature of a pricing was done by Chen et al. \cite{CHEN08} for  envy free pricing of products over a graph with metric substitutability.

As before, agent $i$ seeks bundle $S_i$ and has valuation $v_i$.

\subsection{Construction of Graph H}

For each $i\in \{1,\ldots,m\}$, define
$$A(i) = \{1 \leq j \leq m| S_i \subseteq S_j \mbox{\rm\ and\ }  v_j < v_i \}.$$
The reason we consider only requests with lower valuations is that when picking $i$'s price only these allocations are at risk.
Given price $p$ for agent $i$, all requests in $A(i)$ with valuation $<p$ cannot be allocated at any price.

For each agent $i$ , define an ordering $\pi_i$ on
$A(i)$ in non-decreasing order of valuation. {\sl I.e.}, for each pair  $j,k$ such that $1 \leq j \leq k \leq n_i$, where $n_i = |A(i)|$,
the valuations must be ordered,  $v_{{\pi_i}(j)} \leq v_{{\pi_i}(k)}$ (ties are broken arbitrarily).

We construct an undirected vertex-weighted graph $H$ as follows. For each $i \in V$ we associate $n_i + 1$ weighted vertices in $H$.
These vertices constitute the $T(i)$ component, the vertices of which are $\{i_1,i_2,...,i_{n_i+1}\}$. The set of all $H$ vertices is
defined as $\bigcup_{i \in V}T_i$. The weight of each vertex in $T(i)$ is:
\begin{eqnarray*}
    w(i_1) &=& v_{{\pi_i}(1)} \\
 w(i_2) &=& v_{{\pi_i}(2)} - v_{{\pi_i}(1)} \\
  &\vdots & \\
 w(i_n) &=& v_{{\pi_i}(n_i)} - v_{{\pi_i}(n_i-1)}  \\
 w(i_{n_i+1}) &=& v_i - v_{{\pi_i}(n_i)}
\end{eqnarray*}

By definition of $A(i)$ and $\pi_i$, all weights are non-negative and $$\sum_{j\in T(i)}w(j)=v_i.$$
For each $T(i)$ component we connect the various vertices of $T(i)$ to components $T(j)$ such that $j \in A(i)$
(connecting a vertex $i_k \in T(i)$ to a component $T(j)$ means connecting $i_k$ to all vertices of $T(j)$ by edges) as follows.
Vertex $i_{n_i+1}$ is connected to all components $T(j)$ for $j \in A(i)$. Vertex $i_k$ is connected to all components $T(\pi_i(j))$
for each $j$ such that $1\leq j < k$. For instance, vertex $i_2$ is connected to component $T(\pi_i(1))$, vertex $i_3$ is connected
to components $T(\pi_i(1))$ and $T(\pi_i(2))$, and so on. As $i$ is not contained in $A(i)$ there can't be self edges.
It is easy to see that for any $a<b$, $i_b$ is connected to each component that $i_a$ is connected to (and maybe to some additional components).

\begin{figure} [hb]
 \centering
 \scalebox{0.35}{\input{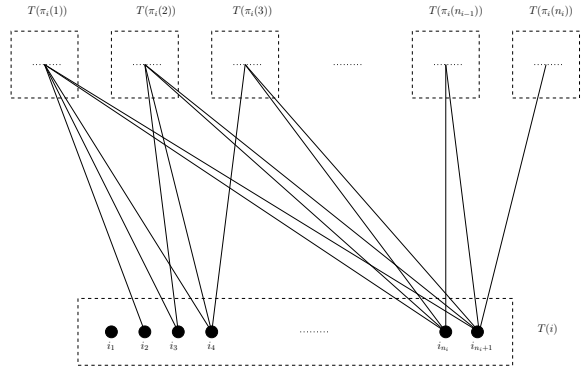}}
 \caption{Construction of $H$}
 \label{fig5}
\end{figure}

\begin{figure} [ht]
 \centering
 \scalebox{0.4}{\input{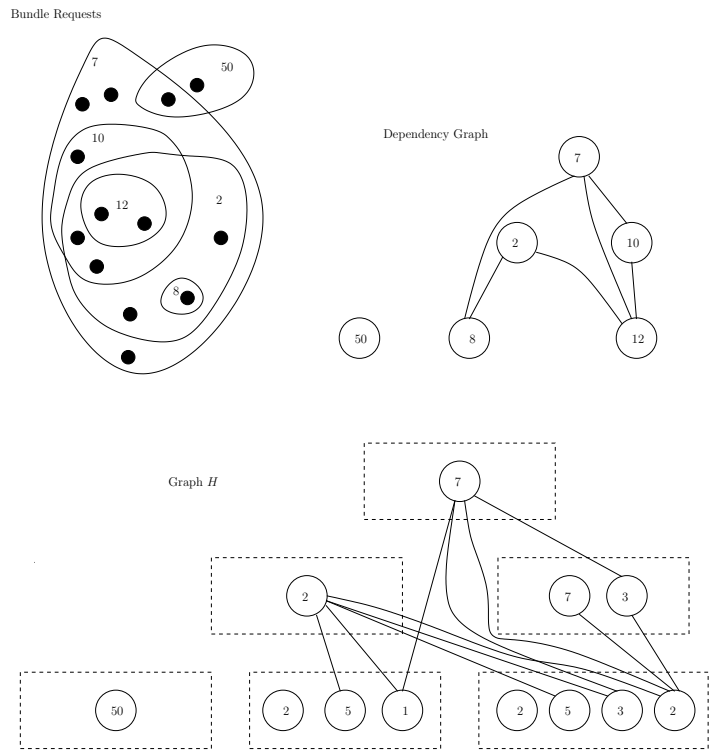}}
 \caption{Step by step example of turning pricing problem into the graph $H$.
  In the bundle requests, each agent would like to buy a set of products (the black balls) as long as its price is less than her valuation
 (the numbers in the bundle requests are valuations).
 $A$ can be seen as a dependency graph where there is a vertical edge from each request $i$ up to the requests of $A(i)$
 (note there is no edge between $2$ and $7$ since $2 \leq7$).
 At the last step the dependency graph is translated into the graph $H$ as defined. }
 \label{fig4}
\end{figure}

See Figure \ref{fig5} for the exact construction of edges from a component $t(i)$. Figure \ref{fig4} shows an example of
transforming a pricing problem into the graph $H$.

\begin{lem} \label{p:IS}
The value of the maximum weighted independent set on $H$ is equal to the revenue obtained from the optimal envy free pricing of the original problem.
\end{lem}
\begin{proof}
When picking vertices for the maximal independent set instance, in every component $T(i)$ one must choose vertices whose sum equals the valuation of agent $j$, where $j\in A(i)$ or $j=i$. A component that none of its vertices were picked means that agent $i$ was
not allocated her set. The construction of $H$ ensures that the pricing is envy-free. We prove that a maximal revenue
envy-free allcation/pricing can be translated into an independent set in $H$ and that a maximal independent set in $H$ can be translated into a revenue maximizing envy free allocation/pricing.

$(envy free  \Rightarrow IS)$ We show how to construct an independent set solution in
$H$ from the optimal allocation/pricing of the original pricing instance.
It is easy to see that the price $p_i$ is equal to one of the valuations $v_j$ for $j$ such that $j \in A(i)$ or $j=i$ (otherwise the prices can be increased).
We will pick vertices in $T(i)$ to achieve price of $p_i$. Let us assume that $p_i = v_j$ for some $j$ such that $j \in A(i)\cup \{i\}$.
We will pick all vertices of $i_k \in T(i)$ such that $k \leq \pi(j)$. By construction of $T(i)$ our pick gives an accumulated
value of $v_j = p_i$. As we have a valid pricing we can assume that $\forall i,j : S_i  \subset S_j \Rightarrow p_i \leq p_j$.
Let us assume by contradiction that our pick is not a valid independent set. It follows that there are two vertices $i_k$ and $j_m$
such that $j \in A(i)$ and there is an edge between them. Since edges are drawn from $i_k$ to all components that represent requests
that have lower valuation than $\sum_{1 \leq t < k} w(i_t)$ we get that $p_j$ must be less than $p_i$ ($p_i \geq \sum_{1 \leq t < k} w(i_t)$).
This contradicts our assumption.

$(IS \Rightarrow envy free )$ Assuming we have an optimal independent set solution in $H$. By $H$ construction,
in each component $T(i)$ any node $i_k$ is connected to all neighbors of $i_m$ for $m<k$.
Therefore the vertices set that are picked as part of the independent set in each component $T(i)$ is of the form $\{ i_k | k \leq i^{max}\}$.
We transform the independent set solution into a pricing as follows:
\begin{itemize}
  \item Agent $i$ such that none of $T(i)$'s vertices were picked receives nothing.
  \item Agent $i$ such that the vertices $\{ i_k | k \leq i^{max}\}$ were picked in $T(i)$ receives $S_i $ at price $\sum_{k \leq i^{max}}{w(i_k)}$
\end{itemize}

Assume that the pricing is not envy-free and we have requests $i,j$ such that $S_i \subset S_j$ and $p_i > p_j$.
By the construction of $H$ we can assume that $p_i$ and $p_j$ are equal to $v_{i'}$ and $v_{j'}$
such that $i' \in A(i)$ and $j' \in A(j)$. We picked from $T(i)$ vertices
$\{ i_k : 1 \leq k \leq i^{max} \} $ such that $\sum_{1 \leq k \leq i^{max}} w(i_k) = v_{i'}$. The same goes for $T(j)$.

Let's inspect the vertices in $T(j)$ that should have been picked in order to make $p_j \geq p_i$.

Define $\mathcal{J}$ as the minimal set of vertices in $T(j)$ of the form $\{j_k | j^{max} < k \leq t\}$ such that $\sum_{1 \leq k \leq t} w(j_k) \geq p_i$.
The vertices of $\mathcal{J}$ has outgoing edges (from $i$ to $A(i)$) into components $T(k)$ such that $k \in A(i)$
(requests that are superset of $j$ and $i$ and have lower valuation than $i$).

The vertices reachable from $\mathcal{J}$ by the outgoing edges are reachable also by $i_{i^{max}}$.
Hence we are guaranteed that they are not picked. Also we know that all vertices that have edges into $\mathcal{J}$
were not picked as picking one of them would have prevented picking any of $T(j)$'s vertices.

Therefore there is no reason for the independent set not to pick also the vertices of $\mathcal{J}$ and increase the independent set value.
This contradicts the maximality of the independent set solution.
\end{proof}

\begin{lem} \label{p:COMP}
$H$ is a comparability graph
\end{lem}
\begin{proof}
Let us direct the edges of $H$. In case of edge from node of $T(i)$ to node of $T(j)$  the direction will be from $i$ to $j$ if $j \in A(i)$.

This orientation assignment results a directed graph with transitivity: if we have directed edge from $i_\alpha$ to $j_\beta$ and from $j_\beta$ to $k_\gamma$, we show that there must be directed edge from $i_\alpha$ to $k_\gamma$.

Note that since $k \in A(j)$ and $ j \in A(i)$ then $ k \in A(i)$ as $A(j) \subset A(i)$. The fact that there is directed edge from $j_\beta$ to $T(k)$ means that $v_j > v_k$, therefore in $\pi_i$ order it will also be higher (And to the right in Figure \ref{fig5}).

Since $i_\alpha$ is connected to all components $T(\pi_i(l))$ such that $1\leq l < \alpha$ and $i_\alpha$ is also connected to $T(j)$, $i_\alpha$ must be connected to $T(k)$ as well (as $\pi_i^{-1}(k) < \pi_i^{-1}(j)$). Clearly $i_\alpha$ is connected to each node of $T(k)$ including $k_\gamma$.

We've shown that the edges of the graph can be oriented so that the transitivity property is maintained.
Therefore the graph is a comparability graph.
\end{proof}

A graph is said to be perfect if the chromatic number (the least number of colors needed to color the graph)
of every induced subgraph equals the clique number of that subgraph. Showing that $H$ is a comparability graph
implies the following corollary:

\begin{cor}
$H$ is a perfect graph.
\end{cor}

We know that the maximal weighted independent set can be solved in polynomial time on perfect graphs \cite{GRO81}.
By Lemma \ref{p:IS} and Lemma \ref{p:COMP} we conclude that finding the optimal envy free
allocation/pricing in the most general single minded setting can be done in polynomial time. This completes the proof of Theorem \ref{p:THM_EF}.

\section{Hardness of multi envy-free allocation/pricing}

In this section we show the following hardness results:
\begin{itemize} \item
The problem of deciding whether a certain pricing assignment is multi envy free is \textit{coNP}-hard. \item Maximizing revenue from single minded agents subject to
multi envy free pricing is \textit{APX}-hard.
\end{itemize}

\begin{thm}
The problem of deciding whether a certain pricing assignment is multi envy free is \textit{coNP-hard}.
\end{thm}
\begin{proof}
We show a polynomial reduction from \textit{VERTEX-COVER(k)} to our decision problem.
Assume that there is an algorithm $\A$ that confirm the envy-freeness of a given subset pricing.
The NP-hard problem of \textit{VERTEX-COVER(k)} can be reduced to this problem.
The building of subset pricing from \textit{VERTEX-COVER(k)} instance is as follows:
\begin{itemize}
  \item Each edge turns into an item.
  \item Each vertex turns into a set with price 1.
  \item Give the price $k-1$ to the set of all items.
\end{itemize}

(Note that this is a limited supply setting where each item is chosen by 3 subsets at most.)

Clearly $\A$ confirms that this instance is not legal if and only if there is a vertex cover to the \textit{VERTEX-COVER} instance of size $k$.

Since \textit{VERTEX-COVER} is \textit{NP-hard} this imply that The problem of deciding weather a certain pricing is envy-free is \textit{coNP-hard}.

\end{proof}

Note that even though deciding whether a pricing is multi envy free is hard, finding such a pricing can be
approximated. Balcan et al. ~\cite{Bal08} showed $O(\log m+\log n)$-approximation for arbitrary bundle valuations and
unlimited supply using single fixed pricing which is basically pricing all bundles at the same price. Such a pricing is multi envy free pricing as well.

\begin{thm} Maximizing revenue from single minded agents subject to
multi envy free pricing is APX-hard, even when all agents are interested in at most two items.
\end{thm}
\begin{proof}
We show that finding the optimal multi envy free pricing is APX-hard by a reduction from MAX-2SAT. Given a MAX-2SAT
instance we build multi envy free allocation instance as follows.

Let's denote $C$ as the number of clauses in the SAT and $C^{(v)}$ the number of clauses containing variable $v$.

Items in the allocation problems are:
\begin{itemize}
  \item For each literal we have an item to sell. Thus for each variable there are two items, one for the variable and one for its negation.
\end{itemize}

Agents in the allocation are:
\begin{itemize}
  \item For each literal of variable $v$ we have two sets of $10C^{(v)}$ agents each. For one set all valuations are $3$
  and for the other all valuations are $2$.
  \item For each variable $v$ we have set of $15C^{(v)}$ agents that wish to buy the variable's item and its negation's with valuation $5$.
  \item For each disjunction clause there is a single agent that wishes to buy both literals of the clause with valuation $5$.
\end{itemize}

We show that there is a pricing with revenue at least $314C + k$ if and only if there is a solution to the 2SAT
 instance that satisfies at least $k$  clauses.

Let us prove the easy direction. Assume that there is a solution to the MAX-2SAT problem that satisfy at least $k$ clauses.
Assign a price of $3$ to each request for a literal that is set to true in the solution. Assign a price of $2$ to each request for literals
that are set to false in the solution. Assign a price of $5$ for each clause request. Each variable request is priced at $5$.
This gives a valid multi envy free pricing and we can verify that the revenue of it is $314C + k$.

For the other direction, let $p$  be a pricing with maximum revenue, and assume the the revenue is at least $314C + k$.
The optimal way to price variable and literal agents for variable $v$ is be pricing one literal at $2$ and the other at $3$,
in that way the variable requests for both literals is priced at $5$ and the revenue from the variable and literals agents is $155C^{(v)}$.
By the optimality of $p$, since variable and literals agents are always more profitable than the clause agents,
all variables must be priced in this manner. Each clause can be priced at $4$ if both its literals are priced at $2$,
or by $5$ if one (at least) of its literals is priced at $3$. In total this means that from pricing allocation of revenue $314C + k$
defines a natural assignment to the MAX-2SAT problem by making literals priced at $3$ to be true.

Because the maximum 2-SAT solution satisfy at least $\frac{1}{4}$ of the clauses, we are seek for a case where $k \geq \frac{c}{4}$.
Some straightforward calculations shows that a $\frac{1256 + \eta}{1257}$ approximation of the multi envy free pricing/allocation
problem would yield an $\eta$-approximation to the MAX-2-SAT. This proves that multi envy free pricing/allocation problem is
NP-hard to approximate to within $\frac{1256 + \eta}{1257}$, where $\eta = .943$ is the approximation hardness constant for
MAX-2SAT problem shown in \cite{KHOT07}.
\end{proof}

\section{The Highway Problem}

The highway problem is the vertex problem pricing in the special case where
the vertices are numbered $1,...,n$ and each agent is interested in an interval $[i,j]$.

\subsection{Multi Envy Free Hardness Results}

\begin{thm} \label{thm:mefhw}
Multi envy free allocation/pricing for the highway problem is in \textit{NP}.
\end{thm}
\begin{proof}
We give a polynomial time algorithm that verifies that a given pricing and allocation is envy free.
The algorithm builds a directed graph over the same nodes of the highway, where for each segment $\mathcal{I}$ there is an edge
for any allocated request $i$ that contains $\mathcal{I}$, with weight $v_i$. Then the algorithm computes the shortest path for any
allocated request's segment in order to find irregularities.
See Algorithm \ref{p:ALG1}.

\begin{algorithm*}[ht]
\begin{enumerate}
  \item  Create directed graph $G$  where for each allocated bundle of price $p$ there are directed edges between the first node
  and all other nodes in the bundle with weight $p$ ( see Figure \ref{fig1})
  \item  For each allocated bundle of price $q$ and for each unallocated bundle with valuation $q$ do:
    \begin{itemize}
      \item Compute the shortest path in $G$ between the first and last nodes of the bundle
      \item  If  $q$ is higher than the shortest path, return false (the allocation/pricing is not multi envy free)
    \end{itemize}
  \item return true
\end{enumerate}
\caption{Verifying a given allocation/pricing on the highway to be multi envy free}
 \label{p:ALG1}
\end{algorithm*}

\begin{figure} [h]
 \centering
 \scalebox{0.55}{\input{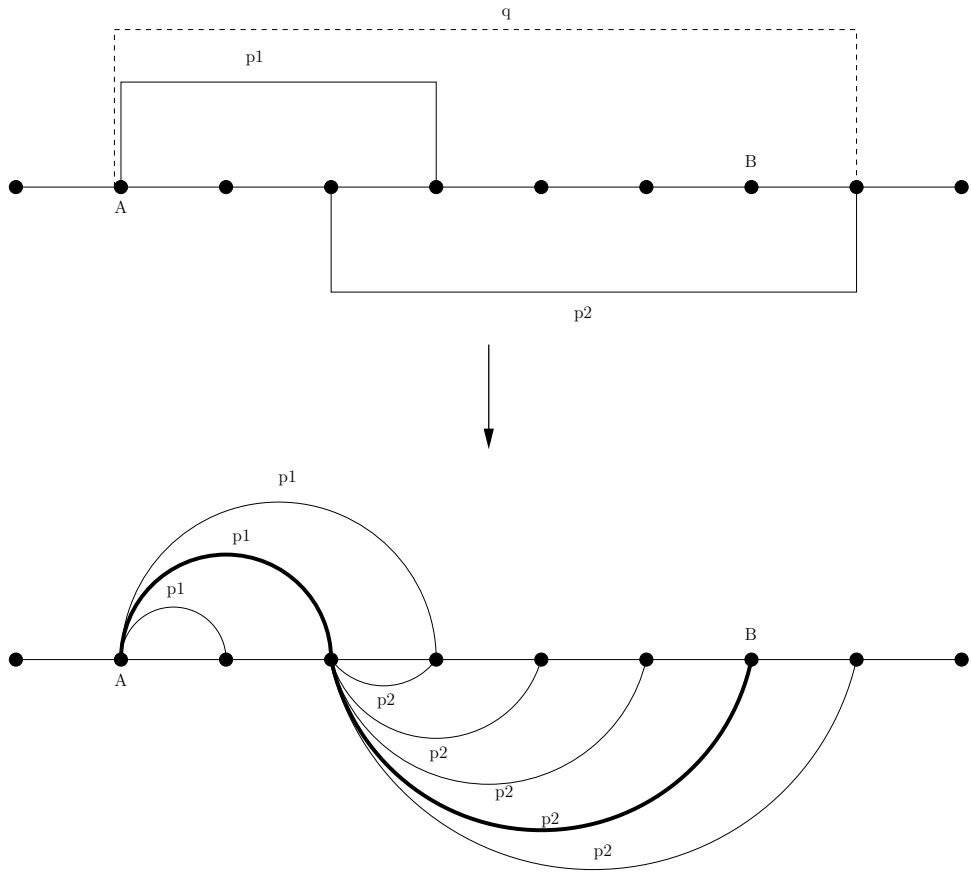}}
 \caption{Translating pricing assignment into graph $G$ and finding path that covers the interval $[A,B]$ on a price lower than $q$. If such path exists, the request for interval $[A,B]$ is envious.}
 \label{fig1}
\end{figure}

Every path of weight $w$ can be translated into a set of allocated bundles of total price $w$ and vice versa.
Therefore is there is no path shorter than $q$ there is no set of requests that has total total price lower than $q$ and the agent is not envious. In the other direction it is easy to see that if the algorithm finds a path which is shorter than $q$ then this path can be translated into a set of agents that get the bundle at a lower price than $q$.
\end{proof}

\begin{thm} \label{thm:revmaxmefhw}
The problem of finding the revenue maximizing multi envy free solution for the highway problem is NP-hard.
\end{thm}
\begin{proof}
We show a polynomial time reduction from \textit{PARTITION}, similar to the reduction of item pricing highway problem shown in ~\cite{Kry06}.
The input to the partition instance is a multiset of weights $I=\{w_i\}$.
For weight $w_i$ we construct a weight component on the highway $\W_i$ which consist of three agents interested in item $i$:
\begin{itemize}
  \item  Request at price $w_i$
  \item  Another request at price $w_i$
  \item  Request at price $2w_i$
\end{itemize}
In addition another two agents are interested in purchasing all the items. Both of them with valuation $\frac{3}{2}\sum{w_i}$. See Figure \ref{fig2}.

\begin{figure} [h]
 \centering
 \scalebox{0.55}{\input{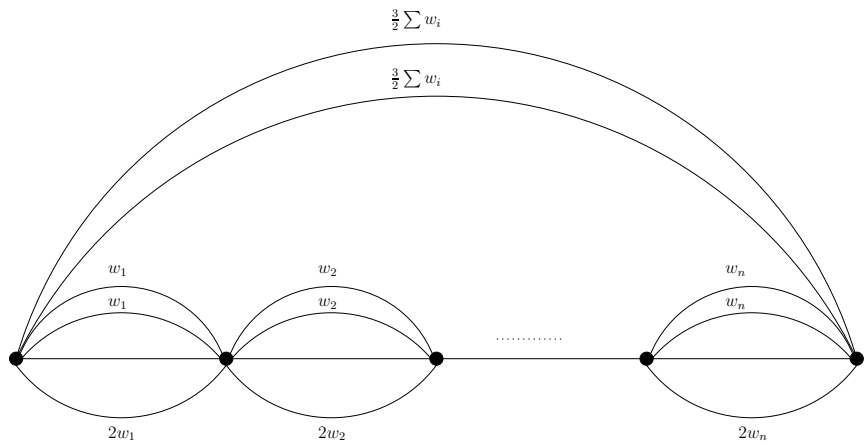}}
 \caption{Construction of multi envy free revenue maximization problem from \textit{PARTITION}}
 \label{fig2}
\end{figure}

The weight obtained from component  $\W_i$  can be $2w_i$, or $3w_i$.
$3w_i$ is achieved by pricing the item at $w_i$ and accepting all agents with price $w_i$ each.
$2w_i$ is achieved by pricing the item at $2w_i$ and accepting only the third agent.
In order to profit from the full valuations of the two agents interested in all items we need that the total price of all items be no more than  $\frac{3}{2}\sum{w_i}$.

It can be argued that the maximum revenue is earned when there is a partition between components, when some of them earn $3w_i$ and some $2w_i$ and the two agents interested in all items pay their full valuations. There is a pricing that reach revenue of $\frac{9}{2}$ if and only if there is a partition of $I$ into $S$ and $I\setminus S$ such that the sum of weights in both is equal.
\end{proof}

\subsection{$O(1)$ Edge Capacities, Multi Envy Free Highway Allocation/Pricing}

In the problem on a path with limited edge capacities, each edge $e \in E$ can accommodate
no more than $c_e$ allocated requests. Let $C = \max{c_e}$. Somewhat inspired by what's done in ~\cite{Gri06} we show that we can solve this problem in time $O(m^{2C}B^{2C^2}n)$ by finding a longest path in an acyclic digraph.

Here is a useful definition of winner multi envy free allocation/pricing and a lemma that shows how to transform winner
multi envy free allocation/pricing to multi envy free allocation/pricing.
\begin{defn}
An allocation/pricing is \textit{winner multi envy free} if
   for any winning agent $i$, its set $S_i$
   is not a subset of a union of other sets (of winning agents) for which the sum of the prices is strictly less than the price of the set $S_i$.

\end{defn}

\begin{lem} \label{p:LOCAL}
Assume we have a winner multi envy free allocation/pricing for a subset pricing instance over a highway with revenue $R$. A multi envy free allocation/pricing can be computed in time $O(m^2)$ with revenue $\geq R$.
\end{lem}
\begin{proof}
Given an allocation/pricing that is winner multi envy free, one can convert it into an allocation/pricing that achieves at least the same revenue and is multi envy free.

For each agent $i$ such that $a_i=\emptyset$ compute the cheapest (by pricing) collection $B$ of winning agents such
that $S_i \subset \cup_{j\in P} a_j$. This can be done by using shortest path algorithm on winning agents prices
in the same way as Algorithm \ref{p:ALG1}.
 If $v_i(S_i) \geq \sum_{j\in B} p(a_j)$ then agent $i$ obeys the condition required for unallocated agents in multi envy free allocations.

If $v_i(S_i) < \sum_{j\in B} p(a_j)$, then we perform the following steps:
\begin{itemize}
    \item Compute the cheapest (by valuation) collection $C$ of winning agents such that $S_i \subset \cup_{j\in C} a_j$ (this can be done by the same way as before).
    \item For each agent $j$ in $C$, if $p(j) < v_j$, set $p(j) = v_j$ and change the prices
        of any allocated agent $k$ such that
        $S_j \bigcap S_k \neq \emptyset$ from $p(a_k)$ to the accumulated payment amount of the minimal weight (by payment) collection of agents that is superset of $S_k$. $C$ contains the agents that are part of the cheapest set of agents among their path, therefore none of
        the agent prices can exceed their valuation.
    \item If $v_i(S_i) < \sum_{j\in C} p(a_j)$ then we assign agent $i$ the set $S_i$,
        set $p_i(S_i) = \sum_{j\in C} p(a_j)$, and set $a_j=\emptyset$ for $j\in C$. Since each agent $j$ in $C$ was allocated with $p(a_j)=v_j$, making $a_j=\emptyset$ does not make $j$ envious.
\end{itemize}

By doing each replacement we clearly still have a winner multi envy free solution.
In addition, for each $j\in C$ there is no cheaper set of agents for $S_j$ (otherwise these agents would have composed $C$ instead of $j$), Therefore none of the agents in $C$ is envious after the switch.

After $m$ iterations the solution is multi envy free.
\end{proof}

\begin{thm} \label{p:THM_HWY}
For a highway with $n$ elements and $m$ agents with maximal valuation $B$ where the capacities of the edges are $\leq C$, there is a $O(m^{2C}B^{2C^2}n)$ time algorithm for the profit maximization  multi envy free problem
on a path.

\end{thm}
\begin{proof}

We create an $n$-layered digraph D with an additional source $s$ and sink $t$, layers $0$ and
$n + 1$, respectively. There are arcs only between layers that represents neighboring items on the highway. Hence,
in any $s \rightarrow t$ path, there are exactly $n + 2$ nodes.

In each node in layer $e$, corresponding
to item $e$, we store all winning agents $j$ that are accommodated by edge $e$. We store
the total  amounts all these agents spend on all items (network links) in their
 path. Moreover, we store for each pair $<i,j>$ the value of the shortest possible path between first
 edge of $i$ to current edge that accommodates $j$. Basically these values can be thought of as matrix  $(A)_{i,j}$ of size $\leq C^{2}$ that holds in each cell the shortest paths between first
 edge of $i$ to current edge that accommodates $j$, then the  diagonal $(i,i)$  represents the amount spent by $i$ itself.

 Any node $x$ (more precisely, the path $s \rightarrow x$)
in the digraph represents a feasible partial solution. Arcs from node $x$ of layer $e$ to node
$y$ of layer $e + 1$ are only introduced if the path $s \rightarrow y$ represents a feasible extension of the
partial solution represented by the path $s \rightarrow x$.
The weight on an arc that connects a node
of layer $e$ to a node of layer $e + 1$ is equal to the profit earned on edge $e + 1$, that is, the
total amount that the new introduced allocated agents of edge $e + 1$ pay.

Therefore, the weight
of the longest $s \rightarrow t$ path in digraph $D$ is equal to the maximum total profit. Moreover,
the set of winning agents can be reconstructed from the longest $s \rightarrow t$ path.
Algorithm \ref{p:ALG2}
shows a more formal description. The allocated agents in this allocation do not envy each other,
however they can be envied by the losers (the allocation/pricing is \textit{winner multi envy free}).
Lemma \ref{p:LOCAL} shows how to overcome this issue and produce \textit{multi envy free} allocation/pricing.

\begin{algorithm*}[ht]
\begin{itemize}
  \item (construction of digraph $D$) For each edge $e \in E$, we introduce a layer of nodes: Denote by $J^{e}$ the set of agents $j$ with
$e \in S_j$ . Let  $U_k^{e}$ holds all possible (sorted arbitrarily) subsets  of $J^{e}$ of size $k \leq \max \{ c_e, |J^{e}| \} $. Define $\mathcal{H} = \{ 0,1,...,B \}$ where $B$ is the maximal valuation over all agents. Let
$H_k$ be the set of all square matrices over $\mathcal{H}$ of size $\leq k$.
A pair $(U,H) \in U_k^{e} \times H_k $ represents having all agents in $U$ as the only winning agents of $e$. The price agent $i \in U$ pays for $S_i$ is $h_{ii}$ and the value of the shortest possible path that pass through $j \in U$, start at the first arc of $S_i$ and ends at $e$ is $h_{ij}$.
The set of all \textbf{nodes} in $D$ is $\bigcup_{e \in E}\bigcup_{k \leq \max \{ c_e, |J^{e}| \}} U_k^{e} \times H_k$ after we remove:
\begin{enumerate}
  \item Illegal pricing nodes that have $h_{ii} > v_i$ \label{p:nodes1}
  \item Nodes that correspond to the last item of $S_i$ for some agent $i$ and $ \exists j \in J^{e}: h_{ii} > h_{ij}$ (this means that $i$ is envy of a group of agents including $j$ that paid less for $i$'s bundle. \label{p:nodes2}
\end{enumerate}

Draw an \textbf{arc} from node $(U,H) \in U_k^{e} \times H_k $ to $(U',H') \in U_{k'}^{e+1} \times H_{k'} $ for layers $2,...,n$ if the following conditions hold: (As $U$ and $U'$ are sorted, for simplicity, we denote a request as $j \in U$ and $j' \in U'$. When saying that $j' \not \in U'$ the meaning is that all elements of $U'$ does not relate to the same agent as $j \in U$.)
\begin{enumerate}
  \item If agent $j \in U$ and ${e+1} \in S_j $ then $j$ appears also in $U'$. If $e \in S_j$ and $j \not \in U$ then $j' \not \in U'$ \label{p:arc1}
  \item For agent $j \in U$ that also appears as $j' \in U'$ the pricing is consistent, formally: $h_{jj} = {h'}_{j'j'}$
  \item For  pair of agents $i,j \in U$ that both appear also in $U'$ as $i',j'$ respectively, the shortest path from beginning of $i$ through $j$ stays consistent, formally: $h_{ij}=h'_{i'j'}$
  \item For agent $j'\in U'\setminus U$ (meaning its correspondent $j$ does not appear in $U$) and for each $i' \in U'$, $h'_{i'j'}$ should be equal to $h'_{j'j'} + {\min_{d} h_{id}}$ (if $i \not \in U $ then the $\min$ expression is set to zero) - this gives the shortest path computation through each added agent to the allocation
\end{enumerate}
We connect node $s$ to all nodes in layer $1$ and we connect all nodes in layer $n+1 $ to $t$.
For each arch that goes from node $(U,H) \in U_k^{e} \times H_k $ into node $(U',H') \in U_{k'}^{e+1} \times H_{k'} $ we give weight of $\sum_{j'\in U'\setminus U}h'_{j'j'}$.
  \item Compute the longest $s \rightarrow t$ path $\mathcal{P}$ in digraph $D$. The winner set is the set of agents that appear in  $U$ component of some node of $\mathcal{P}$. The pricing of certain agent that appear as $j \in U$ of some node of $\mathcal{P}$ is $h_{jj}$ of the same node.
\end{itemize}

\caption{Finding winner multi envy free pricing given an instance of the limited edge capacities highway problem}
\label{p:ALG2}
\end{algorithm*}

\begin{lem}  \label{p:DYNM}
There is a $O(m^{2C}B^{2C^2}n)$ time algorithm that produces optimal winner multi envy free
allocation/pricing for the profit maximization problem on the highway.
\end{lem}
\begin{proof}
Recall that $C$ is an upper bound on the edge capacities.
Consider path  $\mathcal{P}$ from $s$ to $t$ in $D$. The winner set is the union of all winning agents of nodes of  $\mathcal{P}$.
By the construction of $D$ its nodes on level $e$ can't accommodate more than $c_e$ and agent $i$ can't  get item $e$ that does not belong to $S_i$.
By condition \ref{p:arc1} of the arcs definition an agent can be allocated with her entire bundle or with an empty bundle.
By definition of the nodes set (condition \ref{p:nodes1}) all allocations with higher price than agent's valuation are removed.
By condition \ref{p:nodes2} of the node set no agent can envy other agents in  $\mathcal{P}$.

We showed that  $\mathcal{P}$ is giving a legal allocation/pricing that is  \textit{winner multi envy free}.
Since the total weight of  $\mathcal{P}$ gives the revenue (each winning agent price is summed once, when it first appears)
we get that the heaviest path $\mathcal{P}$ yields an optimal solution.

The size of $D$ is
Each edge in the original graph translated to layer of nodes in $D$. For a layer there are at most $m^{C}$
possible subsets of size $\leq \max \{ c_e, |J^{e}| \} $ which is multiplied by the size of $\bigcup_k{U_k^{e}}$
which is bounded by $B^{C^2}$. Therefore there are at most $m^{C}B^{C^2}$ nodes in a layer.
Each node in a layer has at most $n^{C}B^{C^2}$ edges to nodes in the next layer,
this gives total of $m^{2C}B^{2C^2}$ arcs between two consecutive layers.
This means that there are at most $m^{2C}B^{2C^2}n$ arcs in $D$.
The computation time to find the longest path in $D$ is linear in the number of arcs, since D is acyclic ~\cite{ahujamagnantiorlin93}.

\end{proof}

We continue with the proof of Theorem \ref{p:THM_HWY}. By Lemma \ref{p:DYNM} we can build a winner multi envy free
optimal solution to the problem in time  $O(m^{2C}B^{2C^2}n)$. Then if we use simple algorithm (similar to Algorithm \ref{p:ALG1})
that computes the smallest valuation collection of winners for each envious losing agent, from Lemma \ref{p:LOCAL}
we get an $O(m^{2C}B^{2C^2}n)$ algorithm as required.
\end{proof}

\subsection{FPTAS for Highway Revenue, $O(1)$ Edge Capacities,  Multi Envy Freeness}
We next show how to turn the dynamic programming algorithm into a fully polynomial
time approximation scheme (FPTAS); that is, for any $\epsilon > 0$, we have an algorithm
that computes a solution with profit at least $(1 - \epsilon)$ times the optimum profit, in time
polynomial in the input and $\frac{1}{\epsilon}$. To that end, we just apply the dynamic programming
algorithm on a rounded instance in which the agents' valuations are $b'_j = \lfloor b_j / K \rfloor$ where
$K := ({\epsilon B}/{m(n+1)})$ for $\epsilon > 0$.

We show an FPTAS for the problem of finding an optimal winner multi envy free solution.
By Lemma \ref{p:LOCAL} we also get an FPTAS for the multi envy free problem as well.

Let us denote by $(W,p)$ an allocation of winners ($W$) and prices for bundles ($p$). Let $\prod (W,p)$ denote the revenue of the instance $(W,p)$.
\begin{lem}  \label{p:multi envy free_PTAS_LEM1}
For any winner multi envy free solution of the original $(W,p)$ instance, there is a winner multi envy free solution $(W,p')$ of the rounded instance so that $\prod (W,p) > \frac{1}{K}\prod (W,p') - mn -m$
\end{lem}
\begin{proof}
Let $(W, p)$ be a feasible solution to the original instance with profit $\prod(W, p)$. At first for each bundle we set its price in the rounded instance as $p'_i = \lfloor p_i / K \rfloor$. Now we have a pricing that might be illegal, hence our second phase is to set each illegally priced bundle $i$ (when agent $i$ is envy) to be with price $\sum p'_j$ of the cheapest collection of bundles that cover $S_i$. During the first phase each bundle price is reduced by at most $1$.

During the second phase each bundle price is reduced by at most $n$ (number of items) as the size of the cheapest collection of bundles that cover $S_i$ is bounded by $n$.
The solution $(W,p')$ is feasible to the rounded instance and:
$$\prod(W,p') = \sum_{j \in W}p'_j > \sum_{j \in W}(\frac{p_j}{K}-n-1) \geq \frac{1}{K}\prod(W,p) -mn -m$$
\end{proof}
\begin{lem}  \label{p:multi envy free_PTAS_LEM2}
For any winner multi envy free solution of the rounded $(W',p')$ instance, there is a winner multi envy free solution $(W',\tilde{p})$ of the original instance so that $\prod (W',\tilde{p}) = K \prod (W',p')$.
\end{lem}
\begin{proof}
By multiplying the prices of the rounded instance by $K$ we get prices that are still lower than the valuations (of the original instance), therefore this gives a valid solution for the original instance as required.
\end{proof}
Now we combine the two lemmas to produce an FPTAS.

\begin{thm}
There exists an FPTAS for revenue maximization winner multi envy free highway problem with constant edge capacities.
\end{thm}
\begin{proof}
Let $(W,p)$ be the optimal solution for the original instance. By rounding the valuations $b'_j = \lfloor b_j / K \rfloor$ we can compute optimal solution to the rounded instance $(W',p')$. By the process described in Lemma \ref{p:multi envy free_PTAS_LEM2} we construct solution $(W',\tilde{p})$. By two previous lemmas this solution approximation to the optimal solution is
\begin{eqnarray*}\prod (W', \tilde{p}) &=& K \prod (W',p')\\ &>& K (\frac{1}{K}\prod (W,p') - mn -m )\\ &=&
\prod (W,p') - \epsilon B \frac{m(n+1)}{m(n+1)}\end{eqnarray*}

Ergo,  $\prod (W', \tilde{p}) \geq (1-\epsilon ) \prod (W,p)$.
\end{proof}

Since Lemma \ref{p:LOCAL} shows a polynomial algorithm that transform a winner multi envy free solution in the highway to a multi envy free solution, we get the following corollary:

\begin{thm} \label{thm:main}
There exist an FPTAS for the revenue maximization multi envy free highway problem with constant edge capacities.
\end{thm}

\subsection{$O(1)$ Edge Capacities on Envy Free Subset Pricing on Highway}
The problem can be solved in polynomial time by building an acyclic graph similar to the one in previous section but simpler. The number of possible prices is equal to the number of valuations (rather than $B$ for multi envy free problem) as each price equals to some request valuation. In each level there are nodes for each valid subset of allocated agents accommodated by the segment and valid pricing assignment for these agents. The restriction on the graph is that two requests sharing a segment must not envy each other. By finding longest path on the polynomial size DAG the problem can be solved in polynomial time.

To summarize:
\begin{thm} \label{thm:main2}
There exist polynomial time algorithm for the revenue maximization envy free highway problem with constant edge capacities.
\end{thm} \ref{sec:hw}.

\bibliographystyle{plain}
\bibliography{Xbib}

\end{document}